\documentclass[preprint,prd,aps,showpacs,preprintnumbers,amsmath,amssymb]{revtex4-1}
\usepackage{graphics,epsfig,subfigure}
\usepackage{diagbox}
\usepackage[usenames]{color}
\usepackage{color}
\usepackage{graphicx}
\usepackage{amsfonts}
\usepackage{indentfirst}
\usepackage{booktabs}
\usepackage{hyperref}
\hypersetup{hypertex=ture,backref=true,colorlinks=ture,linkcolor=blue,anchorcolor=blue,citecolor=blue}
\usepackage{float}
\usepackage{multirow}

\begin{document}
\renewcommand{\baselinestretch}{1.3}
\newcommand\beq{\begin{equation}}
\newcommand\eeq{\end{equation}}
\newcommand\beqn{\begin{eqnarray}}
\newcommand\eeqn{\end{eqnarray}}
\newcommand\nn{\nonumber}
\newcommand\fc{\frac}
\newcommand\lt{\left}
\newcommand\rt{\right}
\newcommand\pt{\partial}

\title{\Large \bf Frozen boson stars in an infinite tower of higher-derivative gravity}
\author{Tian-Xiang Ma, Yong-Qiang Wang\footnote{yqwang@lzu.edu.cn, corresponding author
}
}

\affiliation{ $^{1}$Key Laboratory of Quantum Theory and Applications of MoE, Lanzhou Center for Theoretical Physics, Lanzhou University, Lanzhou 730000, China\\
	$^{2}$Key Laboratory of Theoretical Physics of Gansu Province, Institute of Theoretical Physics $\&$ Research Center of Gravitation, Lanzhou University, Lanzhou 730000, China\\
    $^{3}$School of Physical Science and Technology, Lanzhou University, Lanzhou 730000, China}

\begin{abstract}
  In this paper, we present a solution for a five-dimensional boson star under gravity with infinite tower of higher curvature corrections. We discover that when the coupling constant exceeds a certain threshold, an alternative configuration emerges, distinct from the conventional five-dimensional boson star. This new structure is characterized by a broader frequency range, with its minimum value approaching zero. At a truncation of $n=2$ for the correction order, the solution and its scalar curvature diverge as the frequency approaches zero. However, as the order of higher curvature corrections increases, the singularity at the center vanishes, resulting in a globally regular solution. Additionally, as the frequency approaches zero, the scalar field's radial distribution becomes concentrated within the critical radius $r_c$, forming what we term a ``frozen star". Beyond this radius, the metric of the frozen star almost degenerates into that of an extreme  black hole. The solutions for such frozen stars offer a new avenue for exploring the enigmatic interiors of compact celestial bodies, enhancing our understanding of the internal structure of black holes under semi-classical conditions and potentially addressing the series of paradoxes associated with information loss due to singularities and horizons.
\end{abstract}

\maketitle

\section{Introduction}\label{sec1}
Black hole research stands as one of the most enthralling areas within contemporary physics. Within the purview of General Relativity, black holes represent not just a most straightforward and natural solution but also unveil the profound aspects of gravitational theory. The Schwarzschild solution characterizes black holes as regions with event horizons, which aptly describes the spacetime exterior to the black hole. However, the presence of internal singularities~\cite{Hawking:1970zqf, Penrose:1964wq, Senovilla:1998oua} triggers a host of issues, such as infinite energy density at the core and the ultimate fate of particles that plunge into the black hole --- do they vanish at the singularity? These conundrums highlight the limitations of General Relativity, as the existence of singularities is deemed physically untenable. Researchers are skeptical about the actual presence of singularities in nature and anticipate that quantum gravity effects might mitigate them, yet quantum gravity has not yet offered definitive predictions.

The evolution of black hole studies from classical General Relativity to the realms of modified gravity and quantum gravity has been a subject of historical significance and recent heightened interest. Among these models, regular black holes~\cite{Ayon-Beato:1998hmi, Ayon-Beato:2000mjt, Ayon-Beato:2004ywd, Bronnikov:2000vy, Bronnikov:2000yz, Bronnikov:2017sgg, Dymnikova:2004zc, Berej:2006cc, Balart:2014jia, Fan:2016rih, Junior:2023ixh} are particularly striking, as they resolve the central singularity issue inherent in General Relativity. To procure solutions for regular black holes, alterations to gravitational theory or the introduction of exotic matter to circumvent the central singularity are contemplated. The Bardeen black hole~\cite{Bardeen:1968} and Hayward black hole~\cite{Hayward:2005gi} stand as archetypal solutions of regular black holes achieved by incorporating exotic matter, each with a regular core. Moreover, a set of solutions involves compact objects formed through the coupling of scalar fields with Bardeen fields, termed Bardeen-boson stars~\cite{Wang:2023tdz}. Further studies have expanded these solutions to frozen Hayward-boson stars~\cite{Yue:2023sep} and frozen Bardeen-Dirac stars~\cite{Huang:2023fnt}, collectively known as frozen stars. Frozen stars closely resemble quasi-black holes~\cite{Lue:2000qr, Lue:2000nm, Lemos:2003gx,Lemos:2007yh,Lemos:2010te,Bronnikov:2014caa,Bronnikov:2016rwg}, possessing a structure akin to an event horizon which referred to as ``critical horizon". At this position, the spacetime metric components nearly degenerate, approaching singularity-like features, yet the scalar curvature remains globally regular. This property indicates that the event horizon is on the cusp of formation, yet it has not fully formed, rendering the star as if it were ``frozen" in stasis from an external viewpoint. 

Beyond these globally regular solutions, there exists a category of boson star solutions constituted by complex scalar fields, initially introduced by Kaup~\cite{Kaup:1968zz}, with Ruffini and Bonazzola~\cite{Ruffini:1969qy} also deriving similar solutions using real scalar fields. Boson stars are celestial entities that resist gravitational collapse through the uncertainty principle of quantum mechanics, and with the incorporation of self-interaction, their density can attain levels comparable to black holes~\cite{Colpi:1986ye}. These entities are devoid of singularities, thereby preventing physical divergences and aligning more closely with the conventional understanding of compact objects. Lately, researchers have been aspiring to construct compact objects without resorting to exotic matter, proposing various methodologies to tackle these challenges. One such approach is to consider add higher-curvature terms in higher-dimensional spacetime, leading to the discovery of a class of regular black holes~\cite{Bueno:2024dgm}. These black holes do not necessitate the introduction of any exotic or unreasonable matter and are part of quasi-topological gravity~\cite{Oliva:2010eb, Myers:2010ru, Dehghani:2011vu, Ahmed:2017jod, Cisterna:2017umf}. The quasi-normal modes of these regular black holes have been analyzed in Ref.~\cite{Konoplya:2024hfg}, affirming that low-order expansions can serve as satisfactory approximations for the complete regular black hole solutions. Further investigations have been conducted in Refs.
\cite{DiFilippo:2024mwm,Konoplya:2024kih}

The ongoing study of compact objects continues to propel our comprehension of gravity, quantum mechanics, and cosmology. The gravitational waves detected by the LIGO and Virgo collaboration~\cite{LIGOScientific:2016aoc}, along with the observations conducted by ETH on the M87 galaxy and the Milky Way~\cite{EventHorizonTelescope:2019dse, EventHorizonTelescope:2019ths, EventHorizonTelescope:2019ggy, EventHorizonTelescope:2020qrl, EventHorizonTelescope:2022wkp, EventHorizonTelescope:2022urf, EventHorizonTelescope:2022wok, EventHorizonTelescope:2022xqj, EventHorizonTelescope:2022exc}, furnish robust evidence for the existence of black holes. However, current observational data does not confirm the internal structure of these black holes. Concurrently, the investigation of regular black holes intimates that General Relativity might need to be superseded by a more comprehensive theory to explicate the physical phenomena in regions of high density. In this paper, we introduce a complex scalar field under the gravity with infinite tower of higher-curvature corrections in five-dimensional spacetime, yielding frozen star solutions. Moreover, we ascertain that, from the view of a distant observer, the spacetime metrics of these frozen stars align perfectly with those of extreme black holes, offering a potential avenue to probe the internal structure of black holes. These studies present novel pathways for exploring the quantum properties of spacetime.

The paper is organized as follows. In Sec.~\ref{sec2}, we construct a model of a frozen boson star incorporating higher-curvature corrections. \ref{sec3} is dedicated to the determination of the boundary conditions required for solving the frozen boson star. In Sec.~\ref{sec4}, we describe the numerical methods employed and present the numerical results obtained within $n=1,2,3,4,\infty$, then there is followed by a discussion of these results. Finally, We summarize the obtained results in Sec.~\ref{sec5}.

\section{The model setup}\label{sec2}
We consider a gravity theory with higher-curvature terms constructed from arbitrary contractions of the Riemann tensor and the metric, with a Lagrangian density $\mathcal{L}(g^{ab},R_{cdef})$. The action can be expressed as

\begin{equation}\label{equ1}
	S=\int\mathrm d^Dx \frac{\sqrt{|g|}}{16\pi G}\left[R+\sum_{n=2}^{n_{\text{max}}}\alpha_n\mathcal{Z}_n\right]+\mathcal{L}_m, 
\end{equation}
 where $g$ is the determinant of the metric tensor $g_{ab}$, $R$ denotes the Ricci scalar, $\mathcal{Z}_n$ represents the Lagrangian density of the nth order, and $\alpha_n$ is the coupling constant corresponding to the order of correction. The detailed form of $\mathcal{Z}_n$ can be found in Ref.~\cite{Bueno:2024dgm,Bueno:2019ycr}. The Lagrangian density for the matter field, denoted by $\mathcal{L}_m=-g^{\mu\nu}\bar{\Phi}_{, \mu}\Phi_{, \nu}-\mu^{2}\bar{\Phi}\Phi $, is also considered, particularly focusing on the case where the matter field is a scalar field.

We consider the ansatz for a static, spherically symmetric spacetime
\begin{equation}\label{equ2}
	\mathrm ds^2=-\sigma(r)^2N(r)\mathrm dt^2+\frac{\mathrm dr^2}{N(r)}+r^2\mathrm d\Omega_{D-2}^2,
\end{equation}
where $N(r)$ and $\sigma(r)$ are two undetermined functions. The ansatz of scalar field reads a single real scalar $\phi(r)$ with time-harmonic form
\begin{equation}\label{equ3}
	\Phi=\phi(r)e^{-iwt}.
\end{equation}
Since quasi-topological theory does not exist in $D=4$ spacetime, we will now investigate spherically symmetric solutions in five-dimensional spacetime, $i.e.$ $D=5$, by substituting ansatz (\ref{equ2}) and (\ref{equ3}) into (\ref{equ1}) and varying the action, we obtain the equation of motion
\begin{equation}\label{equ4}
	3[r^{4}h(\psi)]^{\prime}=16\pi Gr^{3}(\mu^{2}\phi^{2}+\frac{w^{2}\phi^{2}}{N\sigma^{2}}+N\phi^{\prime2}),
\end{equation}
\begin{equation}\label{equ5}
	 3r^{2}\sigma'\frac{dh(\psi)}{d\psi}=16\pi Gr^{3}\frac{w^{2}\phi^{2}+N^2\sigma^{2}\phi^{\prime2}}{N^{2}\sigma},
\end{equation}
\begin{equation}\label{equ6}
	\phi''+\left(\frac{3}{r}+\frac{N'}{N}+\frac{\sigma'}{\sigma}\right)\phi'+\left(\frac{\omega^2}{N\sigma^2}-\mu^2\right)\frac{\phi}{N}=0,
\end{equation}
where
\begin{equation}\label{equ7}
	h(\psi)\equiv\psi+\sum_{n=2}^{n_{\max}}\alpha^{n-1}\psi^n ,\quad\psi\equiv \frac{1-N(r)}{r^2} .
\end{equation}
The action (1) is invariant under global $U(1)$ transformation $\Phi\to e^{iU}\Phi$ with a constant $U$, which means that there is a conserved current $J_{\mu}$ for this system
\begin{equation}\label{equ8}
    J^\mu=-i(\Phi^*\partial^\mu\Phi-\Phi\partial^\mu\Phi^*),
\end{equation}
Integrate the timelike component of the conserved current on a spacelike hypersurface $\Sigma$, and we obtain the Noether charge:
\begin{equation}\label{equ9}
	Q=\int_{\Sigma}J_\mu n^\mu dV,
\end{equation}
where, $n^\mu$ is the unit normal vector of $\Sigma$. The ADM mass $M$ can be read from the asymptotic sub-leading behaviourof the metric functions:
\begin{equation}\label{equ10}
	g_{tt}=-\sigma(r)^2N(r)=-1+\frac{8GM}{3\pi r^2}+...\quad.
\end{equation}

In the absence of a matter field, the equations of motion reduce to the results of regular black hole presented in Ref.~\cite{Bueno:2024dgm}
\begin{equation}\label{equ11}
	\frac{d\sigma}{dr}=0 ,\quad\frac{d}{dr}\left[r^{4}h(\psi)\right]=0.
\end{equation}
By solving (\ref{equ11}), we can deduce that $\sigma(r)=1$(required by normalization of the time coordinate at infinity), we have
\begin{equation}\label{equ12}
	h(\psi)=\frac{m}{r^{4}},
\end{equation}
where $m$ is an integration constant which is proportional to the ADM mass $M$. It takes the form
\begin{equation}\label{equ13}
	m=\frac{8GM}{3\pi}.
\end{equation}

For $n=2$, by solving (\ref{equ12}), $N(r)$ should be:
\begin{equation}\label{equ14}
	N(r)= 1-\frac{-r^2+\sqrt{\frac{32\alpha GM}{3\pi}+r^4}}{2\alpha},
\end{equation}
Similarly, for $n=3$ and $n=\infty$ (the expression for $n=4$ is too complicated to present), we have the following expressions for $N(r)$:

For $n=3$:
\begin{equation}\label{equ15}
\begin{split}
    N(r)=1-\frac{1}{6}\left(\frac{2^{2/3}\tilde{N}(r)}{\pi^{1/3}\:\alpha^{2}}
    \:-\frac{2\:r^{2}}{\alpha}\:-\:\frac{4\:(2\:\pi)^{1/3}\:r^{4}}{\tilde{N}(r)}\:\right)
\end{split} 
\end{equation}
with
\begin{equation}
    \begin{split}
        \tilde{N}(r)=&\left(7\:\pi\:r^{6}\:\alpha^{3}\:+\:72G\:M\:r^{2}\:\alpha^{4}\:\right.\\
        &\left.+\:3\:\sqrt{\:r^{4}\:\alpha^{6}\:\left(9\:\pi^{2}\:r^{8}\:+\:112G\:M\:\pi\:r^{4}\:\alpha\:+\:576\:G^{2}\:M^{2}\:\alpha^{2}\:\right)}\:\right)^{1/3},
    \end{split}
\end{equation} 
and for $n=\infty$:
\begin{equation}\label{equ16}
	N(r)= 1-\frac{8GMr^2}{3\pi r^4+8GM\alpha}.
\end{equation}

\section{Boundary Conditions}\label{sec3}
To numerically solve the equation derived in Sec.\ref{sec2}, we need to specify boundary conditions for each function to be determined. By analyzing the asymptotic behavior of the differential equation at infinity, the unknown function must satisfy the boundary condition 
\begin{equation}\label{equ17}
    N(\infty)=1,\quad \sigma(\infty)=1,\quad \phi(\infty)=0.
\end{equation}

By expanding the equation near the origin, we can determine the boundary conditions that the unknown function must satisfy at the origin.
\begin{equation}\label{equ18}
    N(0)=1,\quad \sigma(0)=\sigma_0,\quad \partial_r\phi(0)=0.
\end{equation}

\section{Numerical Results}\label{sec4}
To facilitate numerical computations, we set $4\pi G=1, \mu=1$, then we employ the following scaling transformations to get the following dimensionless variables
\begin{equation}\label{equ19}
    r\to r \rho,\quad \omega\to\omega/\rho,\quad \mu\to\mu/\rho,
\end{equation}
where $\rho$ is a positive constant whose dimension is length, and we let the constant $\rho$ be $1/\mu$. Additionally, we introduce a new radial variable
\begin{equation}\label{equ20}
    x=\frac{r}{1+r},
\end{equation}
through this transformation, we can change the range of the radial coordinate from $\tilde{r}\in[0,\infty)$ to a finite interval $x\in[0,1]$, which facilitates numerical solution. 
We utilize the finite element method to numerically solve the system of differential equations. The integration region $0\leq x\leq1$ is discretized into 1000 grid points. The Newton-Raphson method is employed as our iterative approach. To ensure the accuracy of the computational results, we enforce a relative error criterion of less than $10^{-5}$.

\subsection{$n\leq 2$: Einstein and Gauss-Bonnet gravity}
\begin{figure}[!htbp]
  \begin{center}
      \includegraphics[height=.26\textheight]{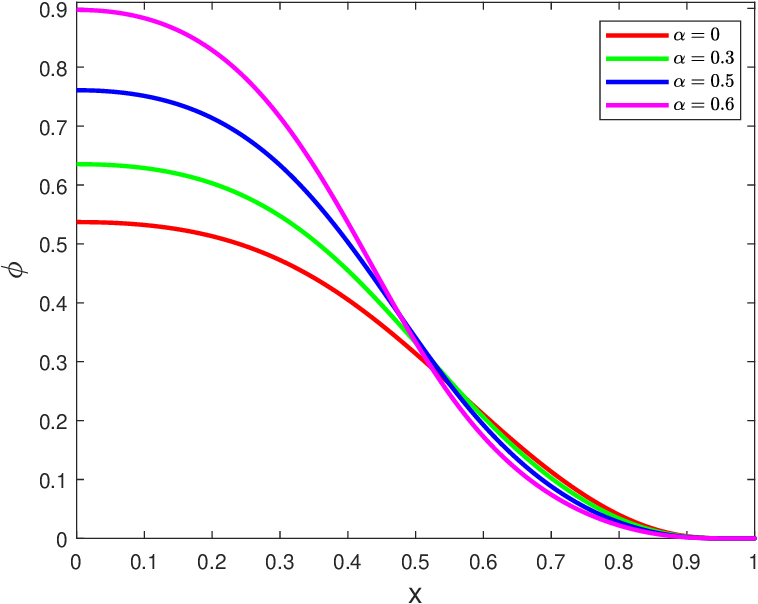}
      \hspace{0.5cm}
      \includegraphics[height=.26\textheight]{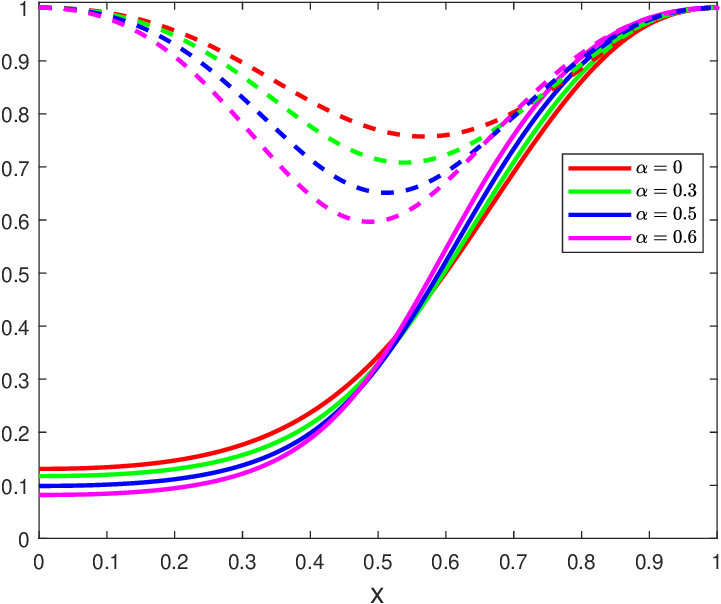}
  \end{center}
  \caption{The scalar field function $\phi$ (left panel), metric component (top right panel) $-g_{tt}$ (solid curve) and $1/g_{rr}$ (dashed curve). All solutions have $\omega=0.96$.}
  \label{field2_alpha}
\end{figure}

\begin{figure}[!htbp]
  \begin{center}
      \includegraphics[height=.26\textheight]{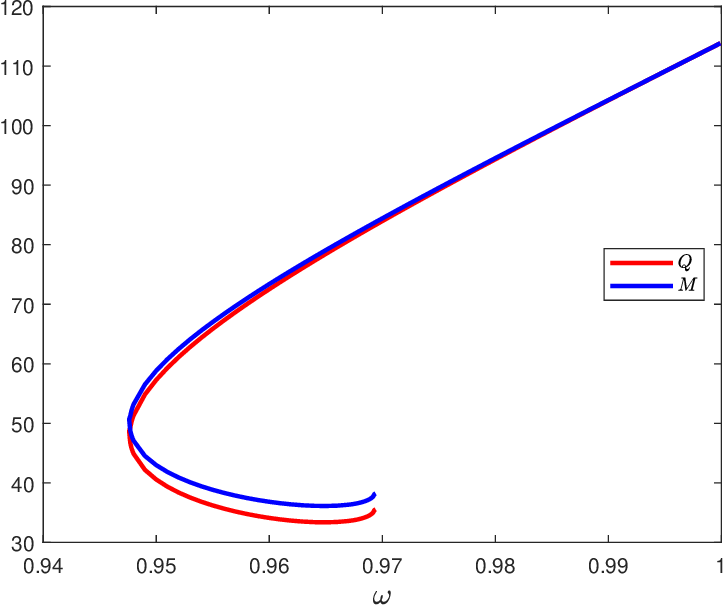}
      \hspace{1cm}\vspace{0.5cm}
      \includegraphics[height=.26\textheight]{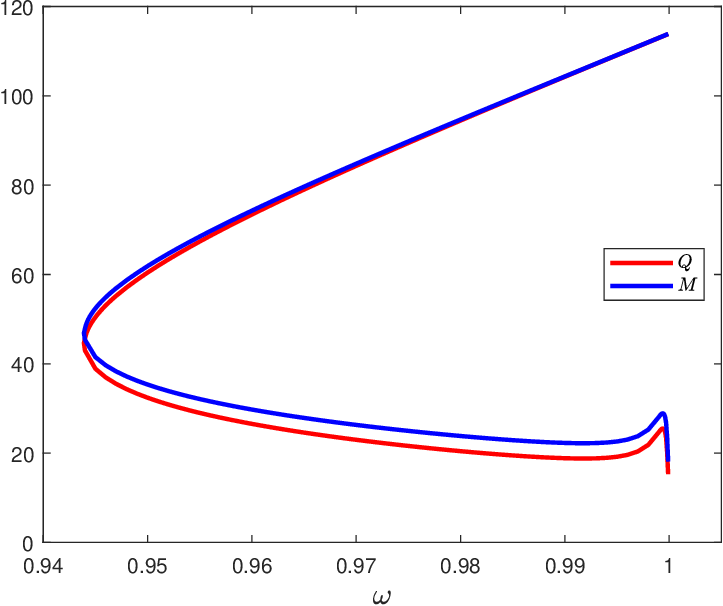}
      \hspace{1cm}\vspace{0.5cm}
      \includegraphics[height=.26\textheight]{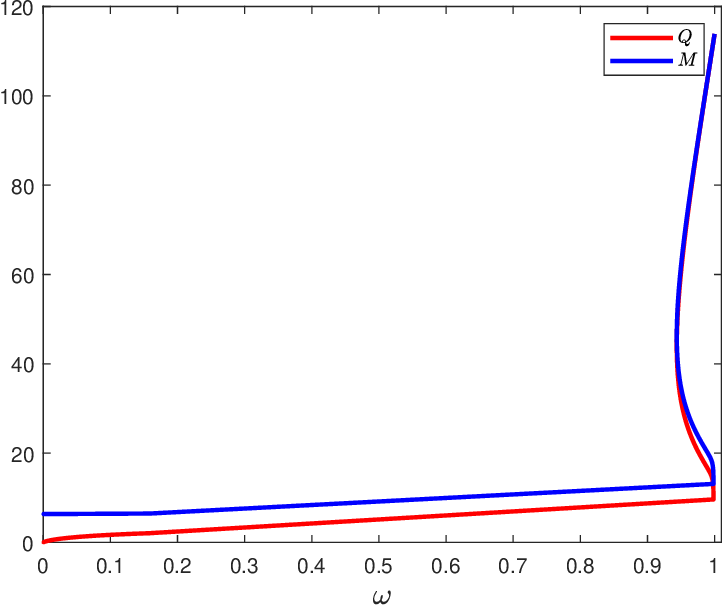}
      \hspace{1cm}\vspace{0.5cm}
      \includegraphics[height=.26\textheight]{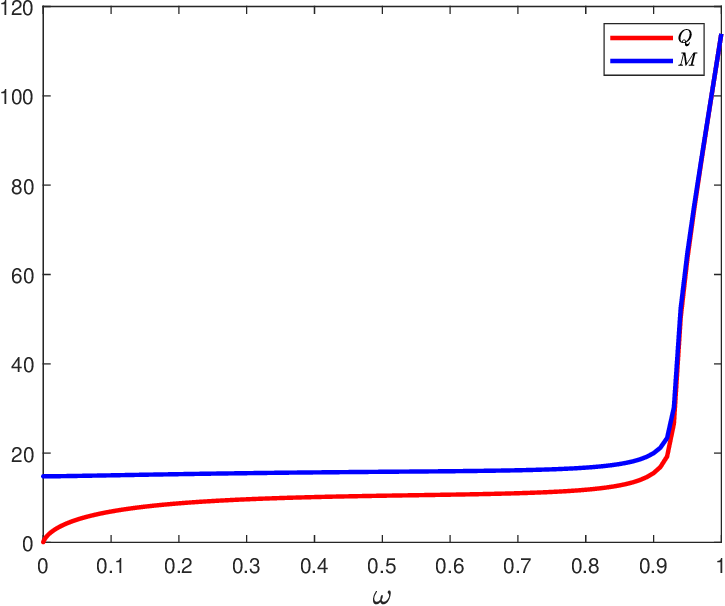}     \hspace{1cm}\vspace{0.5cm}
  \end{center}
  \caption{The Noether charge and ADM mass vs frequency $\omega$ with different $\alpha$. $\alpha=0$ (top left panel), $0.3$ (top right panel), $0.43$ (bottom left panel), $1$ (bottom right panel).}
  \label{qm12}
\end{figure}

\begin{figure}[!htbp]
  \begin{center}
      \includegraphics[height=.26\textheight]{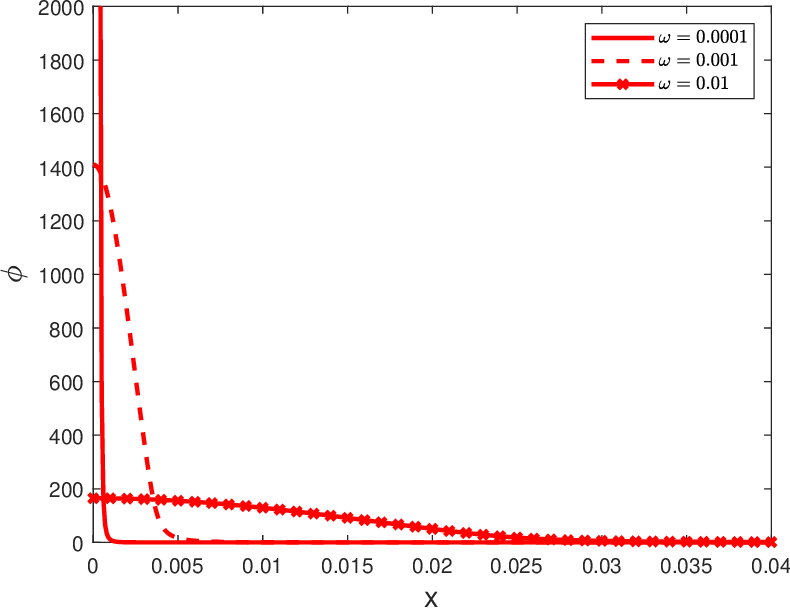}
      \hspace{0.5cm}\vspace{0.5cm}
      \includegraphics[height=.26\textheight]{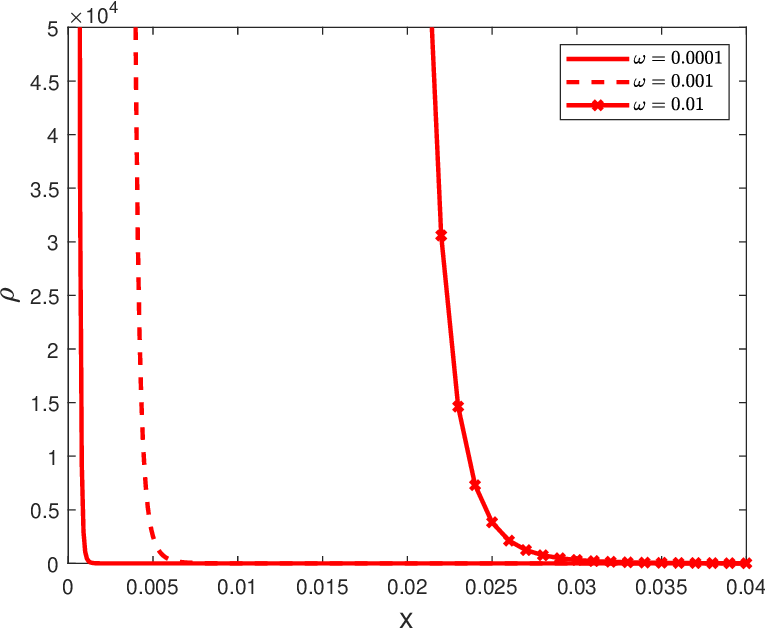}
      \hspace{0.5cm}\vspace{0.5cm}
      \includegraphics[height=.26\textheight]{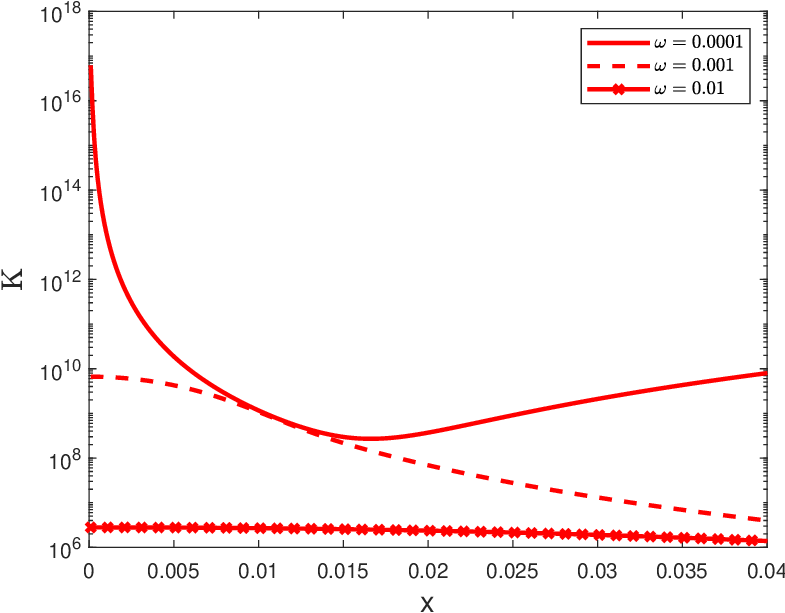}
      \hspace{0.5cm}\vspace{0.5cm}
      \includegraphics[height=.26\textheight]{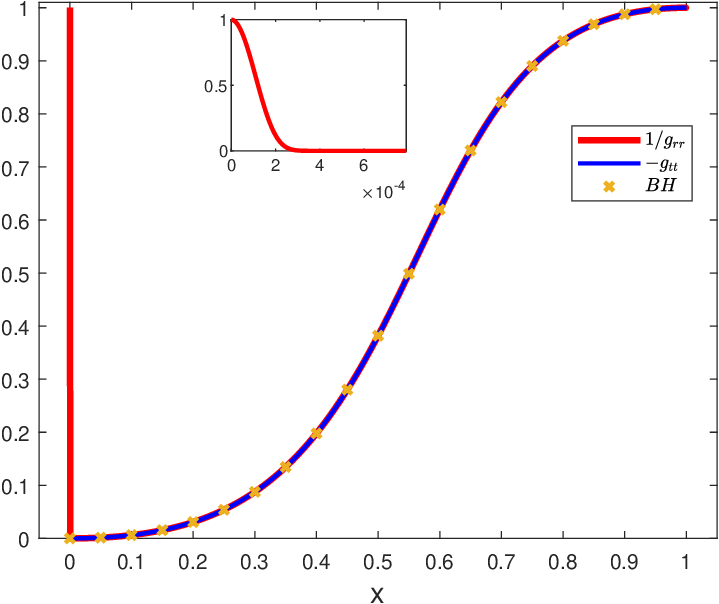}
      \hspace{0.5cm}\vspace{0.5cm}
  \end{center}
  \caption{The matter field (top left panel) and energy density distribution (top right panel) for $n=2$. Kretschmann scalar (bottom left panel) and the comparison between  the boson star with $\omega= 0.0001$ and extreme Gauss-Bonnet black hole (bottom right panel). All solutions have $\alpha=1$.}
  \label{field2_w0}
\end{figure}

\begin{figure}[!htbp]
  \begin{center}
      \includegraphics[height=.26\textheight]{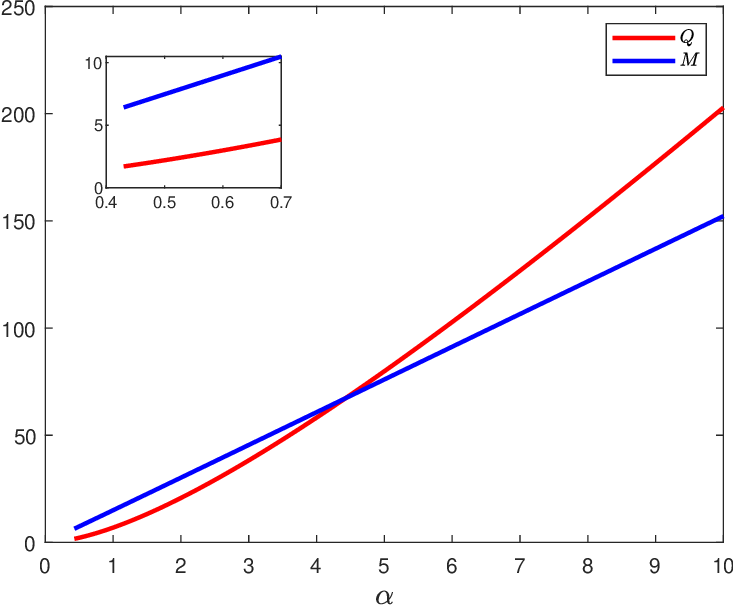}
      \hspace{0.5cm}
      \includegraphics[height=.26\textheight]{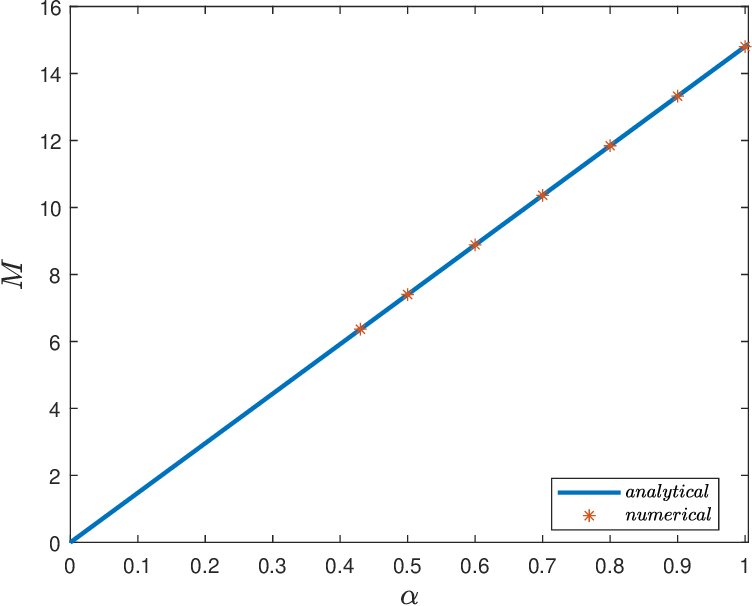}
  \end{center}
  \caption{The Noether charge (ADM mass) vs $\alpha$ with $\omega=0.1$ (left panel). ADM mass $M$ vs $\alpha$ with $\omega\to 0$ (right panel). The blue solid line represents the analytical solutions of 5D extreme black holes, derived from (\ref{equ14}). The orange mark * corresponds to the ADM mass of the numerical results of given $\alpha$ boson stars at $\omega\to 0$. All solutions have $n=2$.}
  \label{n_2compare}
\end{figure}

We have computed the five-dimensional spherically symmetric solutions for $n=1,2,3,4$ and $\infty$, where the matter field comprises solely the ground state scalar field with only gravitational interactions. In this section, we shall revisit the scenarios of $n=1$ (boson star under 5-dimensional Einstein gravity) and $n=2$ (boson star under 5-dimensional Gauss-Bonnet gravity), respectively. In Fig.~\ref{field2_alpha}, we illustrate the impact of different values of $\alpha$ on the field configuration for the case of $n=2$. Notably, when $\alpha=0$, the situation is equivalent to the solution for $n=1$. We observe that as $\alpha$ increases, the peak value of $\phi$ also increases, while the components of the metric, $-g_{tt}$ and $1/g_{rr}$, exhibit decreasing minima with increasing $\alpha$. Fig.~\ref{qm12} illustrates the relationship between the Noether charge and the ADM mass with respect to the frequency $\omega$ corresponding to different values of $\alpha$ for $n=2$. The solution for boson star under 5-dimensional Einstein gravity has been studied in Ref.~\cite{Hartmann:2010pm,Hartmann:2012gw,Blazquez-Salcedo:2019qrz}, where the relationship between $Q$ and $M$ with $\omega$ still manifests as a spiral form, comprising multiple branches. At $\omega \to 1$, we observe $Q \to M$. However, diverging from the four-dimensional case, at $\omega \to 1$ , $Q$ and $M$ are not zero. Moreover, as $\alpha$ increases, the spiral gradually unfolds. After reaching the second branch at $\omega\to 1$, it will extend backward to form a third branch, $\omega$ can be so closed to 0. As $\alpha$ continues to increase, the multiple branches gradually disappear, resulting in a single-valued function. As we traverse along curve, both $Q$ and $M$ monotonically decrease, unlike the four-dimensional boson star, which exhibits a maximum value. We also observe that for solutions where $\omega$ can approach 0, $M$ tends to a constant value as $\omega\to 0$, but $Q\to 0$. Among the solutions mentioned above, it consistently holds that $Q < M$.

In the following discussion, we delve into the essence of these phenomena. The research on five-dimensional boson stars under Gauss-Bonnet gravity has been previously conducted in Ref.~\cite{Hartmann:2013tca,Brihaye:2013zha,Brihaye:2014bqa,Brihaye:2015jja}. In this section, we provide more detailed results. Fig.~\ref{field2_w0} illustrates the field configurations, energy density distributions, and Kretschmann scalar for different values of $\omega$ when $n=2$. An analysis of these numerical results reveals that the Gauss-Bonnet boson star solutions, with $\alpha=1$, exhibit a divergence at the center when $\omega\to 0$. The scalar field, energy density, and Kretschmann scalar all tend towards infinity. As $\omega$ increases slightly, the divergence diminishes, and the matter tends to concentrate within a certain critical radius. This critical radius value expands with an increase in $\omega$. Simultaneously, as the critical radius value increases with $\omega$, the amplitude of the matter field decreases accordingly.

We also compared the $n=2$ boson star solution of $\omega= 0.0001$ with the extreme black hole solution under Gauss-Bonnet gravity with the same ADM mass. From this, we can infer that beyond the critical radius, the metric components of the extreme black hole are identical to those of the boson star. However, due to the divergence in $1/g_{rr}$, it cannot fully degenerate into an extreme black hole. This leads to the result that in Fig.~\ref{qm12}, even when $\omega$ approaches zero and $Q$ is zero, the ADM mass remains finite.

In the left panel of Fig.~\ref{n_2compare}, we illustrate the relationship between the charge $Q$ and mass $M$ of five-dimensional boson star under Gauss-Bonnet gravity corresponding to various parameters of $\alpha\in[0.43,10]$. It is observed that as $\alpha$ decreases, both $Q$ and $M$ diminish monotonically. However, the rate of decrease for $Q$ is more rapid than that for $M$. An excessively small $\alpha$ leads to the vanishing of this configuration, thereby preventing the frequency from asymptotically approaching zero. The relation between the ADM mass and $\alpha$ of extreme black holes under Gauss-Bonnet gravity can be derived from (\ref{equ14}). We obtain that for an extreme black hole with $n=2$, there should be $M=\frac{3\pi^2}{2}\alpha$ according to (\ref{equ14}). By comparing this analytical expression (blue solid line) with the numerical results (orange mark *) for boson stars corresponding to different values of $\alpha$ at $\omega\to 0$, we find that for $n=2$, the boson star’s $M-\alpha$ relation at $\omega\to 0$ is identical to that of extreme black holes. This further supports the conclusion that the boson star under Gauss-Bonnet gravity depicted in the left panel cannot fully degenerate into a black hole.

\subsection{$n>2$: Higher order correction}

\begin{figure}[!htbp]
  \begin{center}
      \includegraphics[height=.26\textheight]{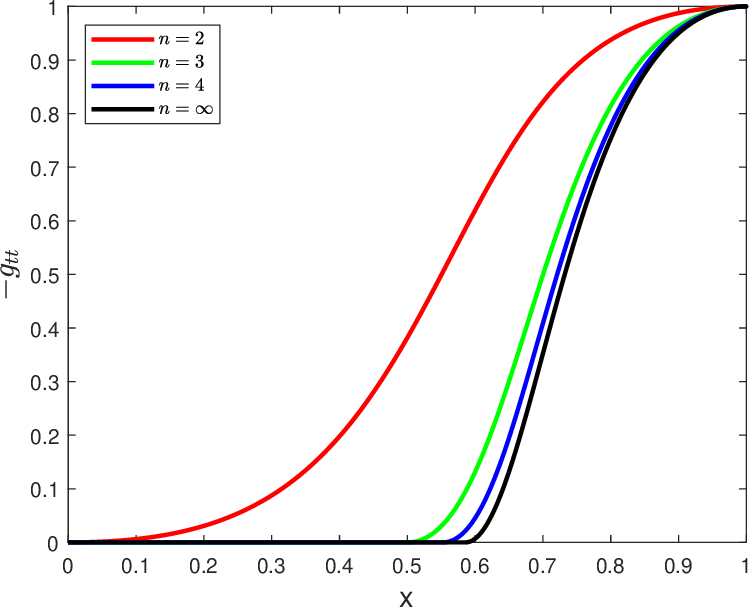}
      \hspace{0.5cm}
      \includegraphics[height=.26\textheight]{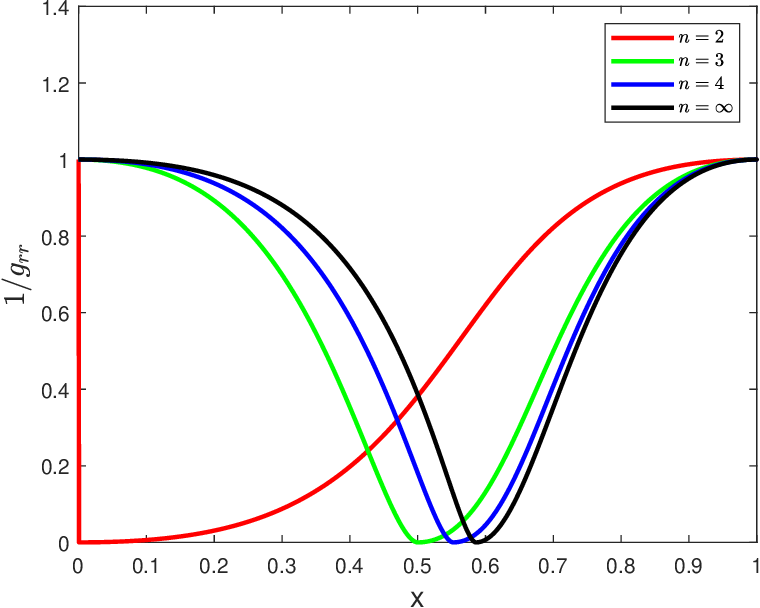}
  \end{center}
  \caption{The metric field function for $n=2, 3, 4$ and $\infty$. All solutions have $\alpha=1$ and $\omega= 0.0001$.}
  \label{metric}
\end{figure}

\begin{figure}[!htbp]
  \begin{center}
      \includegraphics[height=.26\textheight]{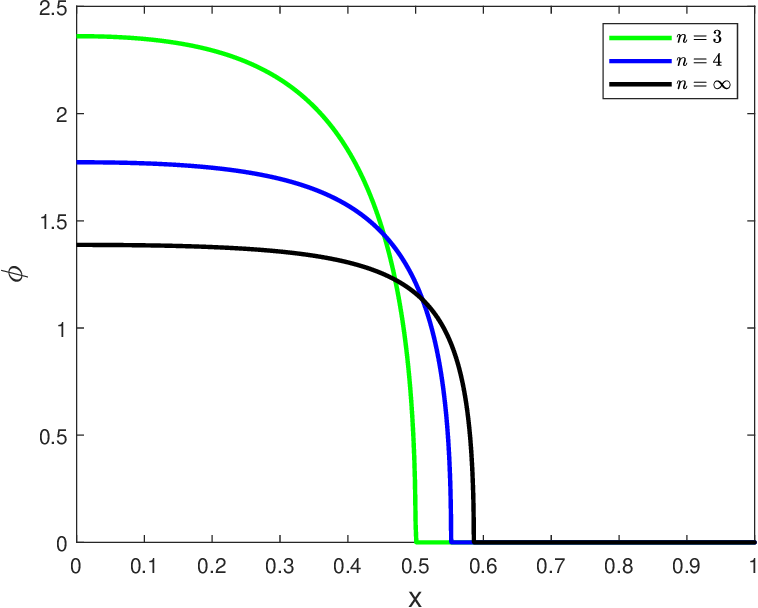}
      \hspace{0.5cm}
      \includegraphics[height=.26\textheight]{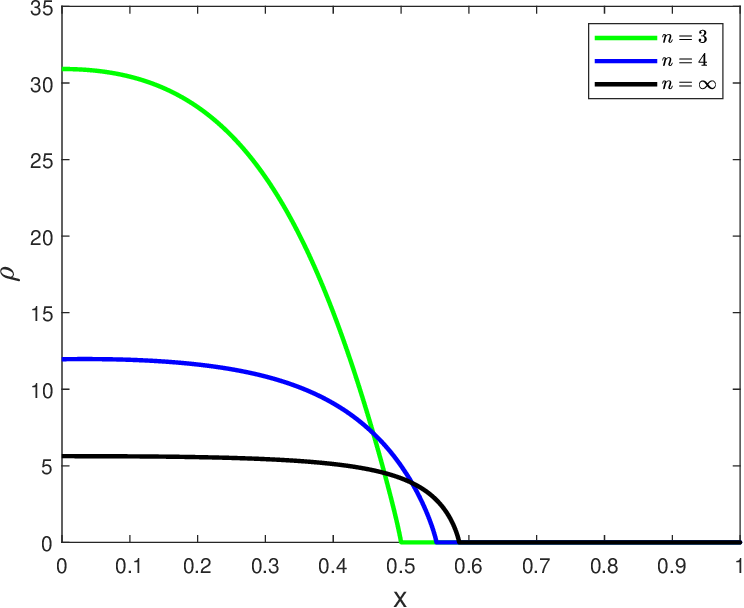}
  \end{center}
  \caption{The matter field (left panel) and energy density distribution (right panel) for $n=3,4$ and $\infty$. All solutions have $\alpha=1$ and $\omega= 0.0001$.}
  \label{field34INF}
\end{figure}

\begin{figure}[!htbp]
  \begin{center}
      \includegraphics[height=.26\textheight]{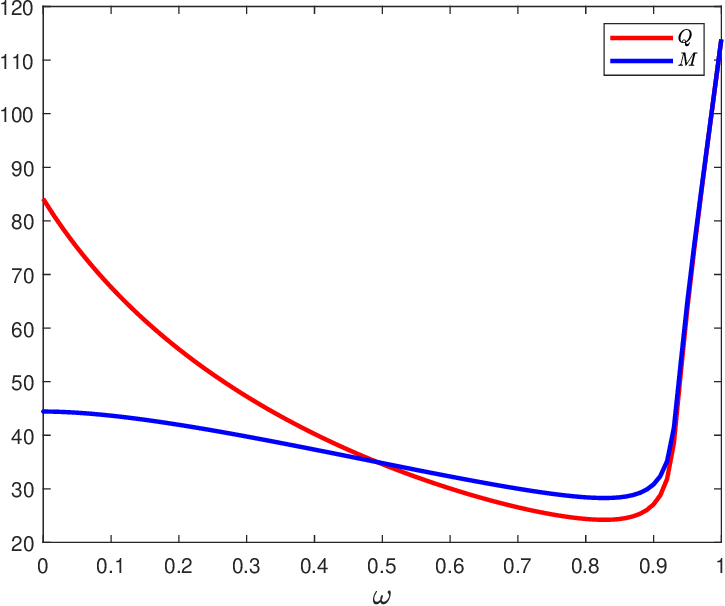}
      \hspace{1cm}\vspace{0.5cm}
      \includegraphics[height=.26\textheight]{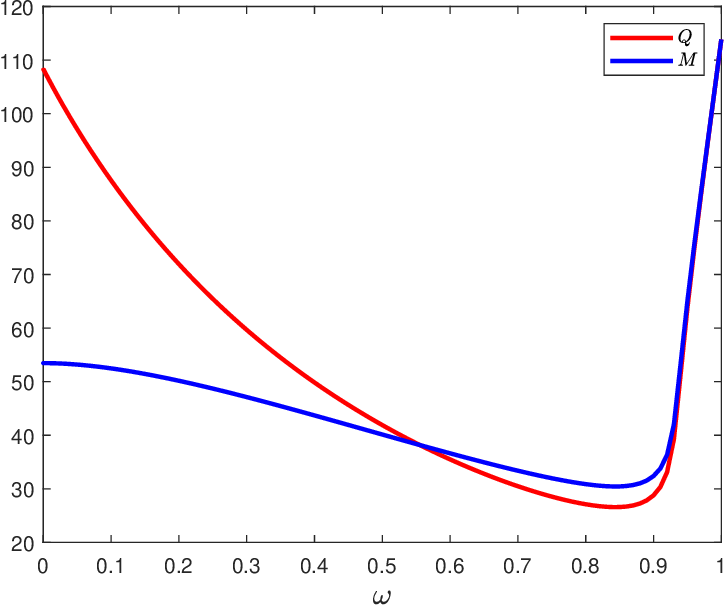}
      \hspace{1cm}\vspace{0.5cm}
      \includegraphics[height=.26\textheight]{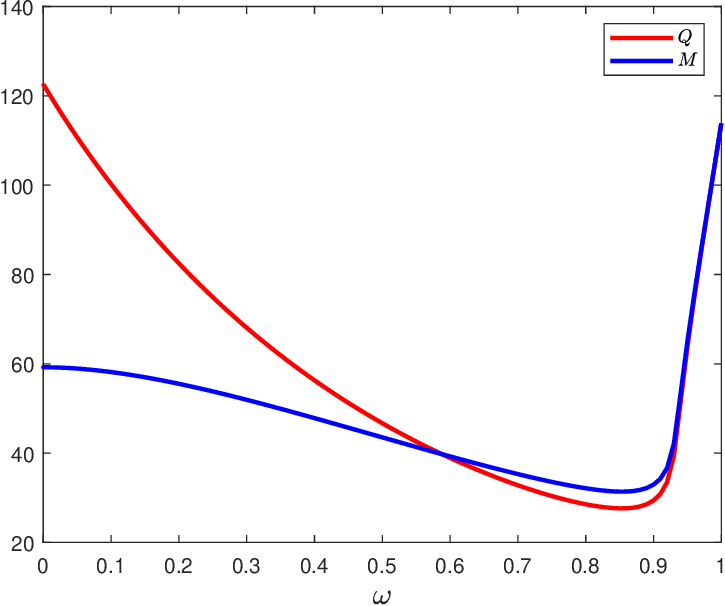}
      \hspace{1cm}\vspace{0.5cm}
  \end{center}
  \caption{The Noether charge and ADM mass vs frequency $\omega$. All solutions have $\alpha=1$ and $\omega= 0.0001$.}
  \label{qm34}
\end{figure}

\begin{figure}[!htbp]
  \begin{center}
      \includegraphics[height=.26\textheight]{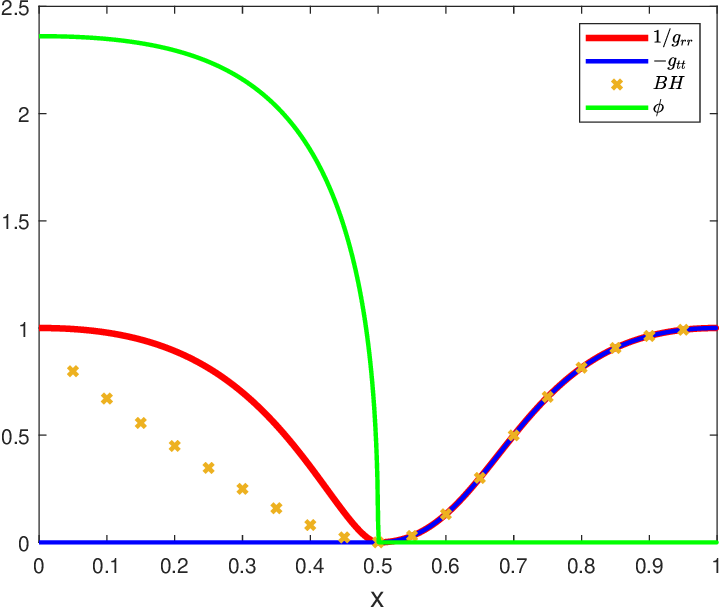}
      \hspace{1cm}\vspace{0.5cm}
      \includegraphics[height=.26\textheight]{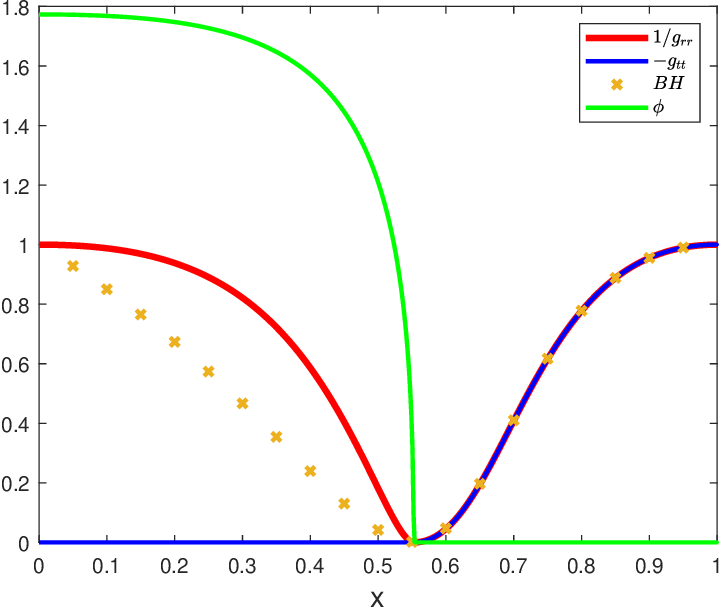}
      \hspace{1cm}\vspace{0.5cm}
      \includegraphics[height=.26\textheight]{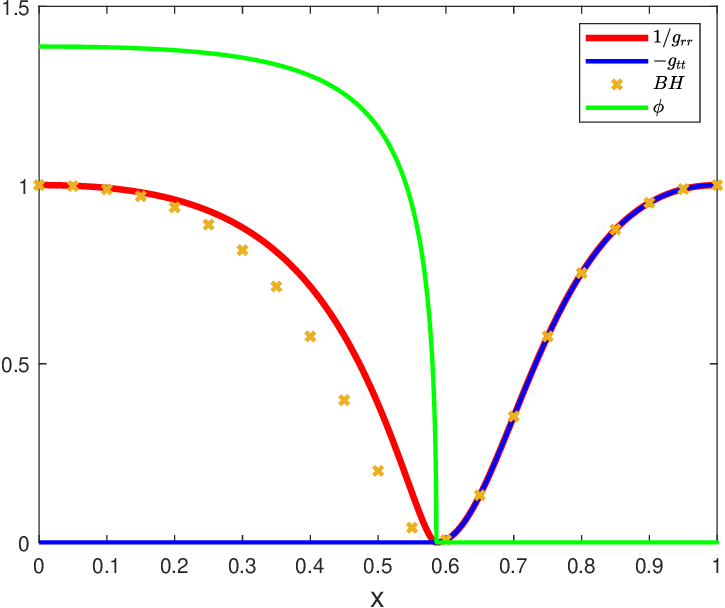}
      \hspace{1cm}\vspace{0.5cm}
  \end{center}
  \caption{The comparison between the frozen star metric and the extreme black hole, both contain the same ADM mass $M$, and both have a value of $\alpha=1$ and $\omega= 0.0001$.}
  \label{extreme}
\end{figure}
In the forthcoming section, we present our numerical results for $n=3,$ $4$ and $\infty$. Fig.~\ref{metric} illustrates the relationship between the metric functions' time component (left panel) and spatial component (right panel) with the radial coordinate, where we set $\omega = 0.0001$. For the case of $n=2$ (red solid line), near $r=0 (x=0)$, the spatial component remains singular, indicating the persistence of spacetime singularities and the absence of a frozen star structure. However, when employing gravity with higher-curvature terms as shown for $n=3$ (green solid line), $n=4$ (blue solid line), and $n=\infty$ (black solid line), the singularity at the center ceases to exist. Moreover, our findings indicate that for these three scenarios, both the metric components $-g_{tt}$ and $ 1/g_{rr} $ approach zero (but they do not get zero) at critical radius $ r_c $, which corresponds to the location of the critical horizon of a frozen star. Additionally, as the value of $n$ increases, so does the value of $r_c$, suggesting that under gravity with higher-curvature terms, the radius of a frozen star expands. 

The conclusions can be readily deduced from Fig.~\ref{field34INF}, which illustrates the function of the matter field and energy density as distributed radially. In contrast to the case with $n=2$, for $n>2$, as $\omega\to 0$, the matter field and energy density are regular at any point in space, becoming increasingly sparse as being farther from center. Beyond the critical horizon $r_c$, the matter field is virtually non-existent. With the increment of $n$, the value of the critical horizon also increases, and concurrently, the amplitude of the scalar field and energy density at the center diminishes. This is consistent with the behavior of the metric components, and the formation of the critical horizon indicates the emergence of a frozen star structure.

Upon incorporating gravitational correction terms, We observed significant changes in the shapes of the $Q$ and $M$ curves compared to $n < 2$ case, as depicted in Fig.~\ref{qm34}. For $n=1$, the $Q(M)$ manifests as a helical shape within the domain of $[0.947, 1)$. When $n=2$, provided that a exceeds a certain threshold, the frequency range can be extended to $(0, 1)$, and the helical shape will no longer emerge. For $n>2$, $Q$ and $M$ initially decrease with $\omega$; however, $Q$ and $M$ begins to increase after the frequency falls below a certain value. As the value of $n$ increases, $Q$ and $M$ at $\omega\to 0$ also increase. 

It is noteworthy that, as concluded in Fig.~\ref{metric}, for the frozen star solution ($n>2$), both  $-g_{tt}$  and $g_{rr}$ tend towards zero at the critical horizon $r_c$, but do not reach zero, marking the position of the critical horizon. We compared this with a black hole possessing the same ADM mass $M$ in (\ref{equ14}--\ref{equ16}) and coupling constant $ \alpha $ as the frozen star solution (Fig.~\ref{extreme}), which happens to be an extreme black hole. Our findings indicate that beyond the critical event horizon, the metric function $-g_{tt} $ and $ 1/g_{rr} $ of a frozen star aligns precisely with that of its corresponding extreme black hole's metric function $ N $, at which point $-g_{tt}$, $1/g_{rr} $, and $N$ are indistinguishable. In other words, it is impossible to differentiate a frozen star from an extreme black hole, external spacetime of frozen star can externally mirror that of an extreme black hole identically. Moreover, since there are no singularities inside and no event horizon, there will be no physically unacceptable divergences or black hole information paradoxes. We consider this to be a potential structure within black holes.

\section{Conclusion}\label{sec5}\
In summary, we have examined the solutions of the five-dimensional spherically symmetric frozen boson star model under gravity with higher-curvature terms. The numerical results indicate that when $n<2$, the frozen star solution does not emerge. When $\alpha>0.43$, the domain of existence for the solution can be $\omega\in (0,1)$. However, in this case, the solution diverges at $\omega\to 0$. Of greater research interest is the case where $n>2$; Here, the matter field and energy are concentrated within the critical horizon $r_c$, and the spacetime is regular globally. Beyond the critical horizon, the spacetime becomes identical with the corresponding extreme black hole which have the same $\alpha$ and $M$ . Consequently, it is impossible for an external observer to distinguish a frozen boson star and an extreme black hole.

Our findings suggest that frozen stars could serve as potential alternatives to black holes, devoid of the singularities and event horizons that lead to divergences and information paradoxes. The resemblance of the spacetime metric of a frozen star to that of an extreme black hole, as observed from infinity, further supports this hypothesis. These results open up new avenues for understanding the intricate structures that may exist within the cosmos, challenging our current perceptions of gravitational interactions and the ultimate fate of massive celestial bodies.

\section*{Acknowledgements}
This work is supported by the National Key Research and Development Program of China (Grant No. 2022YFC2204101 and 2020YFC2201503) and the National Natural Science Foundation of China (Grant No. 12275110  and No. 12247101).

\providecommand{\href}[2]{#2}\begingroup\raggedright

\end{document}